\documentclass[preprint,  amsmath, nofootinbib, showpacs]{revtex4}

\usepackage{graphicx}
\usepackage{amssymb}

\newcommand{\HGeo}{{}_2\!F_1}

\begin{document}
\preprint{AEI--2004--041}
\title{Generating function techniques for loop quantum cosmology}
\author{Daniel Cartin}
\email{cartin@naps.edu}
\affiliation{Naval Academy Preparatory School, 197 Elliot Street, Newport, Rhode Island 02841}

\author{Gaurav Khanna}
\email{gkhanna@umassd.edu}
\affiliation{Physics Department, University of Massachusetts at Dartmouth, North Dartmouth, Massachusetts 02747}
\affiliation{Natural Science Division, Southampton College of Long Island University, Southampton, New York 11968}

\author{Martin Bojowald}
\email{mabo@aei.mpg.de}
\affiliation{Max-Planck-Institut f\"ur Gravitationphysik, Albert-Einstein-Institut, 14476 Potsdam, Germany}

\date{\today}
\begin{abstract} 
Loop quantum cosmology leads to a difference equation for the wave
function of a universe, which in general has solutions changing
rapidly even when the volume changes only slightly. For a
semiclassical regime such small-scale oscillations must be suppressed,
by choosing the parameters of the solution appropriately. For
anisotropic models this is not possible to do numerically by trial and
error; instead, it is shown here for the Bianchi I LRS model how this
can be done analytically, using generating function techniques. Those
techniques can also be applied to more complicated models, and the
results gained allow conclusions about initial value problems for
other systems.

\end{abstract}

\maketitle

\section{Introduction}

One of the two main avenues for current research in quantum gravity is
that of quantum geometry (also known as loop quantum gravity). This is
a theory that leads to space-time that is discrete at Planck-length
scales, giving rise to predictions of quantized areas and volumes (see
\cite{rov04} for reviews). In recent years, these ideas have
been applied to the study of cosmology~\cite{boj-mor04, boj04}. Just as
minisuperspace models reduce the infinitely many degrees of freedom
down to a finite, quantum mechanical model, the basic idea of loop
quantum cosmology (LQC) is to use symmetry reduction to develop
relatively simple models for various space-times. However, many of the
features of LQC are similar to those in the full theory; in
particular, the Hamiltonian constraint is constructed analogously to
the full one \cite{QSDI}, and as a consequence takes the
form of a discrete recursion relation \cite{cosmoIV}.

Loop quantum cosmological models are free of singularities,
irrespective of the type of matter content \cite{Sing}. This is seen
as a general consequence of the quantum evolution equation, which is a
difference equation for the wave function and does not break down
where the classical singularity would be. At the same time, additional
conditions for the wave function can arise as consistency conditions
for solutions of the difference equation whose coefficients can become
zero. These conditions can be interpreted as dynamical initial
conditions \cite{boj01}.  For a vacuum isotropic model
\cite{IsoCosmo}, it turns out that together with the notion of {\it
pre-classicality}~\cite{boj01} a unique LQC solution is picked
out. Pre-classicality requires the coefficients of the wave function
to lack large variations over Planck-scale lengths, and thus allows
the solution to match a semi-classical wave function far away from the
singularity. Because one is dealing with discrete recursion relations,
this means that as the parameter is increased by one, there should not
be a great change in the value of the coefficients. (A precise 
definition of pre-classicality would require more information about 
observables and the physical inner product than is presently available 
\cite{Bohr}.)

The situation is more complicated in anisotropic models, with several
independent directions in minisuperspace along which oscillations need
to be suppressed, and therefore the Hamiltonian constraint becomes a
partial difference equation. We will look in detail at the Bianchi I
LRS model, which has only two independent degrees of freedom and
simplifies thanks to its separable evolution equation.  As we will
see, the generic sequence alternates between positive and negative
values, thus exhibiting huge oscillations. However, with the new
techniques introduced here to LQC it is possible to select out special
boundary values that give pre-classical sequences; from this set of
possibilities, pre-classical wave functions can be built up. Because
these generating function methods can be used to solve generic
(partial) difference equations, they are useful to obtain
pre-classical solutions for other LQC models and their Hamiltonian
constraints.

To study the relation between the quantum theory and its classical
limit, it is of particular interest to construct solutions which in
presumed semiclassical regimes describe wave packets following
classical trajectories at least approximately. This requires a
sufficiently large set of independent pre-classical solutions such
that an initial wave packet can be formed by superposition. That this
is realized is not at all obvious. For instance, the continuum limit
of the difference equation with the corresponding
(DeWitt\footnote{In order to have a regular quantum theory, DeWitt 
introduced the condition that the wave function vanish at points of 
minisuperspace corresponding to classical singularities. 
Posing those initial conditions, however, is not enough to
avoid singularities, and is replaced in loop quantum cosmology by a
different mechanism which depends sensitively on details of a loop
quantization. Still, the dynamical initial conditions mentioned above
take a similar form and emerge automatically from loop quantum
cosmology.}
\cite{DeWitt}) initial condition is not well-posed (solutions to
the Wheeler--DeWitt equation which are zero at the boundaries of
minisuperspace do not suffice to allow reasonable initial data). Even
though there are LQC models which are well-posed in contrast to their
Wheeler--DeWitt analog \cite{Scalar}, it is conceivable that in a
model like that studied here the lack of well-posedness in the
continuum limit precludes the existence of sufficiently many
pre-classical solutions. This indeed turns out to be the case as will
be discussed later together with possible consequences and
interpretations.

The loop quantized Hamiltonian constraint for the particular case of a
Bianchi I LRS space-time used here has already been worked
out~\cite{boj03}. Here we present more systematic methods of finding
solutions with particular properties, which are useful not only for
this space-time, but for any model system to be considered in loop
quantum gravity. In Section \ref{sep-sec}, we discuss in detail how to
use generating functions to obtain information about a sequence with
one parameter. This allows us to tune the magnitude and asymptotic
behavior of the sequence by choosing the appropriate initial
values. By combining two such one-parameter sequences, we can solve
the Bianchi I LRS recursion relation, and pick appropriate
combinations to have the desired properties at large values of the
evolution parameter $n$. Section \ref{full-sec} generalizes the
similar method to solve the full two parameter recursion
relation. However, because of numerical limitations, the actual
sequence is obtained only over a small extent in the parameters $m$
and $n$, but still its properties can be tuned by appropriate choices
of boundary conditions. The general solutions we obtain here are
similar in form to those of Section \ref{sep-sec}. As an application
of the results we discuss the issue of pre-classical solutions and the
semiclassical limit in the Conclusions.

\section{Separable solutions}
\label{sep-sec}

The Bianchi I LRS (locally rotationally symmetric) model is the
simplest anisotropic model, being derived from the Bianchi I model
with symmetry group ${\mathbb R}^3$ by imposing the condition of one
rotational symmetry. There are thus two degrees of freedom left. In
addition, the classical equations of motion of the model can easily be
decoupled, and the quantum Hamiltonian separates. We will first recall
the basic equations of this model from the appendix of \cite{boj03}
and then discuss its separated evolution equations.

\subsection{The Bianchi I LRS model}

The Bianchi I LRS model has two degrees of freedom which, in real
Ashtekar variables, are given by two connection components $(A,c)$ and
the conjugate momenta $(p_A,p_c)$ with sympletic structure given by
$\{A,p_A\}=\frac{1}{2}\gamma\kappa$ and
$\{c,p_c\}=\gamma\kappa$. Here, $\kappa=8\pi G$ is the gravitational
constant and $\gamma$ the Barbero--Immirzi parameter of loop quantum
gravity. The momenta $(p_A,p_c)$ are components of an invariant
densitized triad which determine the scale factors $a_I$ of a Bianchi I
metric by $a_1=\sqrt{|p_c|}=a_2$, $a_3=p_A/\sqrt{|p_c|}$. Thus, the
metric is $d s^2 = |p_c| (dx^2+dy^2)+p_A^2/|p_c| dz^2$ in Cartesian
coordinates and is degenerate for $p_A=0$ or $p_c=0$.

The behavior of $(A,c)$ and $(p_A,p_c)$ as functions of time is
determined by the Hamiltonian constraint
\begin{equation}
 H=-\frac{A(2cp_c+Ap_A)}{\sqrt{|p_c|}}
\end{equation}
which is proportional to $-Ap_A-2cp_c$, leading to decoupled
Hamiltonian equations of motion solved by $p_A\propto\sqrt{\tau}$,
$p_c\propto\tau^2$, $A\propto 1/\sqrt{\tau}$, $c\propto
1/\tau^2$. After introducing a new time coordinate $t:=\tau^{3/2}$, we
obtain $p_A\propto t^{1/3}$ and $p_c\propto t^{4/3}$ and thus scale
factors $a_1=a_2\propto t^{2/3}$, $a_3\propto t^{-1/3}$ whose
exponents $\alpha_1=\alpha_2=2/3$ and $\alpha_3=-1/3$ indeed solve the
Kasner conditions for Bianchi I solutions,
$\sum_I\alpha_I=1=\sum_I\alpha_I^2$ in the special LRS case
$\alpha_1=\alpha_2$.

After quantizing the model with loop techniques \cite{boj03}, the triad
components $p_A$ and $p_c$ become basic operators with discrete
spectra, $\hat{p}_A|m,n\rangle=\frac{1}{4}\gamma\ell_P^2 m|m,n\rangle$
and $\hat{p}_c|m,n\rangle=\frac{1}{2}\gamma\ell_P^2 n|m,n\rangle$. The
wave function is thus supported on a discrete minsuperspace, and as a
solution to the quantized Hamiltonian constraint subject to a
difference equation,
\begin{equation} \label{diff}
d(n) ({\tilde t}_{m-2, n} - 2 {\tilde t}_{m, n} + {\tilde t}_{m+2, n})
+ 2 d_2(m) ({\tilde t}_{m+1, n+1} - {\tilde t}_{m+1, n-1} - {\tilde
t}_{m-1, n+1} + {\tilde t}_{m-1, n-1}) = 0,
\end{equation}
where ${\tilde t}_{m, n}$ are the coefficients of the wave
function (rescaled by a factor of the world volume of each basis state). Also we have
\begin{equation}
d(n) = \biggl \{
\begin{array}{ccr}
0									& \quad	& n = 0
\\
\sqrt{1 + \frac{1}{2n}} - \sqrt{1 - \frac{1}{2n}}	& \quad	& |n| \ge 1
\\
\end{array}
\end{equation}
and
\begin{equation}
d_2 (m) = \biggl \{
\begin{array}{ccr}
0			& \quad	& m = 0
\\
\frac{1}{m}		& \quad	& m \ge 1
\\
\end{array}
\end{equation}
For the remainder of the paper, the parameter $n$ will act as a ``time''
parameter. Defining $t_{m, n} = {\tilde t}_{m+1, n} - {\tilde t}_{m-1,
n}$ ($m \ge 1$), the recursion relation simplifies to
\begin{equation}
\label{diff-eqn}
d(n) [ t_{m+1, n} - t_{m-1, n}]  + 2 d_2 (m) [ t_{m, n+1} - t_{m, n-1}] = 0.
\end{equation}
Note that for typographical simplicity, here and throughout the rest
of the paper the notation $t_{m, n}$ and ${\tilde t}_{m, n}$ has been
reversed from their use in \cite{boj03}. Since ${\tilde t}_{m, n}$ includes the 
volume of each basis state, it follows that $\tilde{t}_{m, n}$ must vanish at the
boundaries, $\tilde{t}_{0,n}=0=\tilde{t}_{m,0}$. Because of its
definition in terms of ${\tilde t}_{m, n}$, the values $t_{0, n}$ are
freely specifiable; the boundary condition $\tilde{t}_{0,n}=0$ is then
used in computing $\tilde{t}$ from $t$ via
$\tilde{t}_{m+1,n}=t_{m,n}+\tilde{t}_{m-1,n}$.  We still must have
$t_{m, 0} = 0$, which will act as boundary conditions for our
sequence.

The continuum limit of the difference equation is obtained at large
triad components (or in the $\gamma\to0$ limit \cite{SemiClass}),
which implies $m, n \gg 1$ such that the difference operators can be
approximated by differentials. One thus arrives at the Wheeler--DeWitt
equation
\begin{equation}
 \frac{1}{2}p_c^{-1}\frac{\partial^2}{\partial
p_A^2}\tilde{\psi}(p_A,p_c)+ 2p_A^{-1}\frac{\partial^2}{\partial
p_A\partial p_c}\tilde{\psi}(p_A,p_c)=0\,,
\end{equation}
where $\tilde \psi$ is the continuous function that interpolates the 
discrete $\tilde t_{m, n}$ for large $m, n$. Note that the difference 
equation (\ref{diff}) for $\tilde{t}_{m, n}$ is of order four, higher than that
of the Wheeler--DeWitt equation, which is always second order. This is
a consequence of the loop quantization and cannot be avoided. It
implies that there are more independent solutions to the discrete
equation than the Wheeler--DeWitt equation has; in fact, not all
the discrete solutions have a continuum limit. Those which do have
such a limit are the pre-classical solutions which do not change
rapidly if the discrete labels $m$ and $n$ are increased. It is known
that, when boundary conditions are not taken into account, there are always
sufficiently many pre-classical solutions for the semiclassical
limit to be achieved \cite{Cons}. In practice, finding those
solutions is usually difficult when also boundary conditions have to
be taken into account.

Solutions of the Wheeler--DeWitt equation can easily be studied after
introducing $\psi(p_A,p_c):=\partial\tilde{\psi}/\partial p_A$ and
separating the resulting differential equation
$p_A\partial\psi/\partial p_A+4p_c\partial\psi/\partial
p_c=0$. Writing $\psi(p_A,p_c)=\alpha(p_A)\beta(p_c)$ we obtain the
conditions $p_A\alpha'=\lambda\alpha$ and $p_c\beta'=-4\lambda\beta$
solved by $\alpha(p_A)\propto p_A^{\lambda}$, $\beta(p_c)\propto
p_c^{-4\lambda}$. Thus, any non-zero solution $\psi$ diverges either
at the boundary $p_A=0$ or at the classical singularity $p_c=0$ unless
$\lambda=0$. For the original wave function $\tilde{\psi}=\int\psi
dp_A$ this implies that solutions regular at the boundary are
available only for $-1\leq\lambda\leq 0$.  DeWitt's condition
$\tilde{\psi}(p_A,0)=\tilde{\psi}(0,p_c)=0$ in this model, which says 
that there is zero probability of finding the universe at a singularity, is thus
ill-posed since not enough separable solutions for constructing given
initial values are available~\cite{boj-mor04}.

\begin{figure}
	\includegraphics[width = 0.8\textwidth]{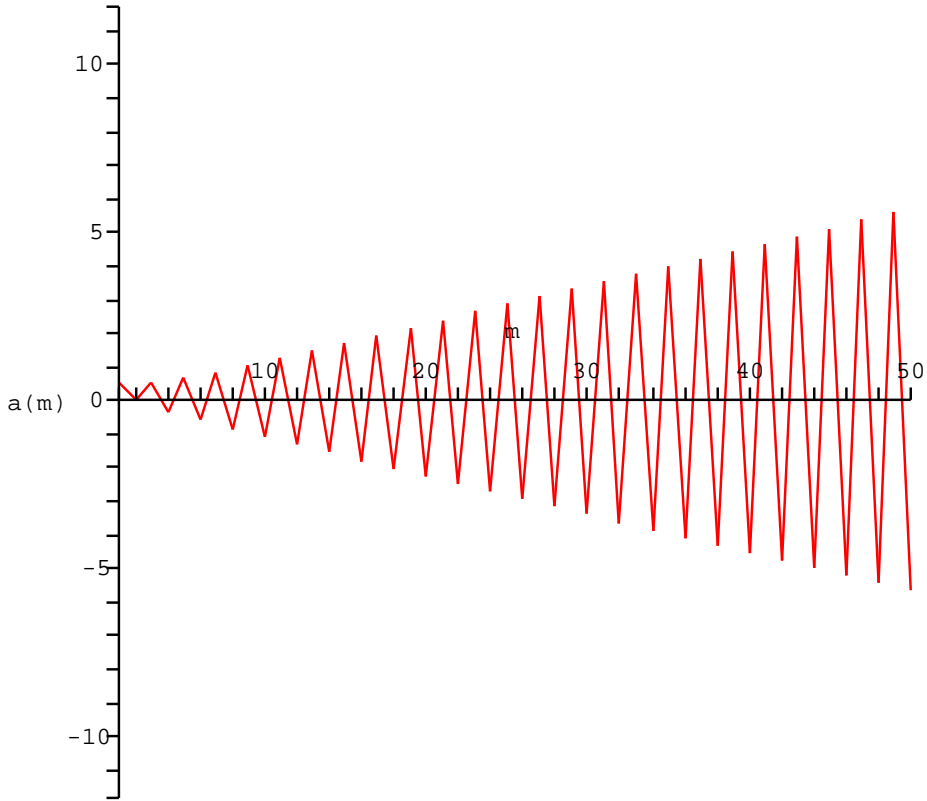}
	\caption{\label{oscill} 
        Oscillatory behavior of a generic
	solution to the recursion relation (\ref{a-eqn}) for the
	sequence $a_m (\lambda = -1)$. The initial values are $a_0 =
	a_1 = 1$.}
\end{figure}

The difference equation $(\ref{diff-eqn})$ for $t_{m, n}$ can be solved similarly by
assuming that the sequence $t_{m, n}$ is separable into two
one-parameter sequences $a_{m}$ and $b_{n}$, i.e. $t_{m,n}=a_mb_n$. This is
possible if $a_{m}$ and $b_{n}$ satisfy
\begin{subequations}
\begin{align}
\label{a-eqn}
a_{m+1} - a_{m-1} & = \frac{2 \lambda}{m} a_{m},
\\
b_{n+1} - b_{n-1} & = -\lambda d(n) b_{n},
\end{align}
\end{subequations}
where $\lambda$ is the separation parameter now for the sequences. Any choice of these two
sequences will solve the original recursion relation; because the
$a_m$ and $b_n$ relations are second-order, the first two values $a_0$
and $a_1$, for example, are enough to describe the rest of the
sequence $a_m$ for a particular $\lambda$. However, as noted before
the order is higher than that of the 1st order separated
Wheeler--DeWitt equations and there are additional solutions with
large oscillations.  For negative $\lambda$, for instance, picking any
two random values for the boundary data will generically give
sequences that alternate between positive and negative values with
magnitudes that increase as the parameter ($m$ or $n$) increases. An
example of such a generic choice is shown in Figure
\ref{oscill}. Because of this alternation, the wave function will not
be smooth at large $n$ and therefore not be pre-classical.

That solutions generically have alternating behavior can be understood
as follows: For large $m$, the right hand side of (\ref{a-eqn}) is
usually small compared to the left hand side such that $a_{m+2}\approx
a_m$. The relation between neighboring values, $a_{m+1}$ and $a_m$, is
determined by the initial values $a_0$ and $a_1$. Together they
determine $a_2=2\lambda a_1+a_0$. For negative $\lambda$, and $a_0=0$,
say, $a_2$ has the opposite sign from $a_1$ which translates to $a_m$
having the opposite sign from $a_{m+1}$ for large $m$ and thus
alternating behavior. If $a_0\not=0$, $a_2$ may have the same sign as
$a_1$, but still generically oscillations will set in later. Only for
positive $\lambda$ is it easy to suppress the oscillations. However,
since $\lambda$ enters the equation for $b_n$ with the sign reversed,
now the $b$-sequence will develop alternating behavior such that the
full solution alternates generically.

We will develop techniques, which can be used analytically or
numerically, to check whether it is possible to choose special initial
values such that oscillations are suppressed. This will fix the
relation between initial values and thus reduce the amount of freedom
in pre-classical solutions. In particular, values like $a_0$ may not
be allowed to vanish. While this does not pose a problem for the
$a$-sequence, for the $b$-sequence we have a different initial value
problem, requiring $b_0=0$ as a consequence of $t_{m,0}=0$. Thus, for
the $b_n$ the only way to suppress oscillations is by restricting to
$\lambda<0$.

\subsection{Generating functions}

Our strategy for choosing the initial values of the sequence $a_m$
will be to work with a generating function\footnote{For a good
reference on generating functions, see Wilf~\cite{wil93}; the related
notion of Z-transforms is covered in, e.g., Oppenheim and
Schafer~\cite{opp-sch75}.}. This is a function of one variable $x$
such that the value $a_m$ is the coefficient of $x^m$ in a Taylor
expansion, i.e.
\begin{equation} \label{expand}
F (x) = \sum_{m=0} ^\infty a_{m} x^m.
\end{equation}
This will allow us to go from a recursion relation for $a_m$ to a
differential equation for $F (x)$. To see how this works, we look at
how derivatives act on the function $x^k F(x)$, for a fixed $k$. We
have
\begin{equation}
\frac{\partial}{\partial x} [x^k F(x)] = \sum_{m=0}^{\infty} (m + k) a_m x^{m + k - 1},
\end{equation}
so we can see that linear functions of $m$ in the recurrence relation
become linear derivatives of the generating function. This makes it
easier to work with $a_m$ instead of $b_n$ because $d_2(m)$, unlike
$d(n)$, is polynomial and the map between recursion relation and
differential equation is more obvious. Techniques for studying the
values of the $b_n$ sequence will be discussed later.

We start with the recursion relation (\ref{a-eqn}) for $a_m$, multiply
it by $m x^{m-1}$ and sum over all values of $m$ for which the
relation is valid (i.e. $m \ge 0$); then the sum is
written as\footnote{We start the sum at $m=0$, so that all
coefficients are of the form $a_{m+k}$, instead of beginning at $m=1$
(with a $a_{m-1}$ term) to avoid placing extra relations on the
coefficients $a_m$. This has the effect of shifting $m \to m+1$.}
\begin{equation}
\sum_{m=0} ^\infty [(m+1) a_{m+2} - 2 \lambda a_{m+1} - (m+1) a_{m}] x^m = 0.
\end{equation}
By mapping instances of $m$ into derivatives, we arrive at the
differential equation for $F(x)$ given by
\begin{equation}
 \frac{d}{dx} \biggl[ \frac{F(x) - a_0}{x} - x F(x) \biggr] -
 2\lambda\frac{F(x) - a_0}{x} = 0
\end{equation}
or
\begin{equation}
 \frac{d}{dx}\biggl[\frac{1-x^2}{x} F(x)\biggr]-
 2\lambda\frac{F(x)}{x}+a_0\frac{1+2\lambda x}{x^2}=0\,.
\end{equation}
As written, this equation has singularities at $x=-1, 0$ and 1, which
makes it problematic to expand around $x=0$ in the Taylor series expansion (\ref{expand}) of the generating function;
however, we are interested not in the solution itself, but in the
relation between the coefficients in its series expansion. Because of
this, there is some freedom to find an equation more tractable to
analysis, so it is natural to define a new function
\begin{equation}
\label{1D-shift}
G(x) =  \frac{F(x) - a_0}{x}.
\end{equation}
Substituting this in to the relation for $F(x)$ gives a simpler
differential equation for $G(x)$,
\begin{equation}
\label{gen-ODE}
\frac{d}{dx} \biggl[ (1 - x^2) G(x) \biggr] - 2\lambda G(x) = a_0.
\end{equation}
By going from $F(x)$ to $G(x)$, we have reduced the number of
singularities by one, but the question now is to relate this new
generating function to the sequence $a_m$. If we take
\begin{equation}
G(x) = \sum_{m=0} ^\infty \alpha_m x^m,
\end{equation}
then the mapping (\ref{1D-shift}) between the generating functions
$F(x)$ and $G(x)$ implies that the two sequences $a_m$ and $\alpha_m$
are related by
\begin{equation}
\label{alph-def}
a_m = \alpha_{m-1}, \qquad {\rm for\ } m \ge 1.
\end{equation}
Thus, simplifying the differential equation has resulted in a ``shift''
upward in the sequences. Once we know the two initial values of the
$\alpha_m$ sequence, $\alpha_0 (= a_1)$ and $\alpha_1 (=a_2)$, then we
can find those of the $a_m$ sequence by using
\begin{equation}
\label{IC-rel}
\alpha_1 - a_0 = 2 \lambda \alpha_0.
\end{equation}
which comes from the recursion relation (\ref{a-eqn}) for $a_m$ in the
particular case $m=1$. Thus, the relation between $\alpha_0$ and
$\alpha_1$ implies one between $a_0$ and $a_1$. The information in
this differential equation is equivalent to that in the recurrence
relation; any series solution of the differential equation
(\ref{gen-ODE}) will solve the relation (\ref{a-eqn}) for $a_m$, after
shifting $\alpha_m$ back to $a_m$ via (\ref{alph-def}).

\subsection{Asymptotic behavior}

The question now is how to avoid alternating oscillatory behavior in the
sequence. We look at the function $(1 - x) G(x)$, which is a function
that generates the {\it differences} between adjacent values of
$\alpha_m$, that is,
\[
(1 - x) G(x) = \alpha_0+\sum_{m=0} ^\infty (\alpha_{m+1} - \alpha_m) x^{m+1}.
\]
If this function has no singularities, then the sequence $\alpha_m$
will converge to a finite value without oscillation. Because of the
$(1 - x^2)$ factor appearing in the differential equation (\ref{gen-ODE}) for
$G(x)$, the function will in general have poles at $x = \pm 1$. Thus
the function is regular for $|x|<1$, and if there is no singularity at
$x=1$, we have $(1-x)G(x)|_{x=1}= \alpha_0+\sum_{m=0}^{\infty}
(\alpha_{m+1}-\alpha_m)= \lim_{m\to\infty}\alpha_m$ such
that the sequence $\alpha_m$ has a finite limit. Moreover, if
there is not a pole at $x=-1$, $\lim_{m\to\infty}(\alpha_{m+1}-\alpha_m)=0$, so alternating
oscillations are suppressed.  This avoids situations such as that seen, for example, in the Taylor
expansion of $(1 + x)^{-1}$, where the coefficients of $x^m$ alternate sign:
\begin{equation}
\frac{1}{1 + x} = 1 - x + x^2 - x^3 + \cdots.
\end{equation}
Obviously, $(1 - x) G(x)$ will have the same behavior at $x=-1$ that
$G(x)$ does; we avoid oscillations by requiring no singularity in the function $G(x)$ at $x=-1$. From this, we get a
relation between the two initial values $\alpha_0$ and $\alpha_1$. The
story is slightly different at $x=1$; $G(x)$ may blow up there, but
$(1 - x) G(x)$ can be finite if, e.g., $G(x)$ has a simple pole at
$x=1$. Two examples of this are seen in the expansions for $(1 - x)^{-1}$,
\[
\frac{1}{1 - x} = 1 + x + x^2 + x^3 + \cdots,
\]
and $\ln(1 - x)$,
\[
 -\ln(1-x) = x + \frac{1}{2} x^2 + \frac{1}{3} x^3 + \cdots.
\]
In the first, the coefficients of the sequence maintain a constant
value; in the second, they converge to zero. We will see that the
latter is exactly the behavior found generically in the solutions
$G(x)$ for negative $\lambda$.

\begin{figure}[htb]
	\includegraphics[width = 0.8\textwidth]{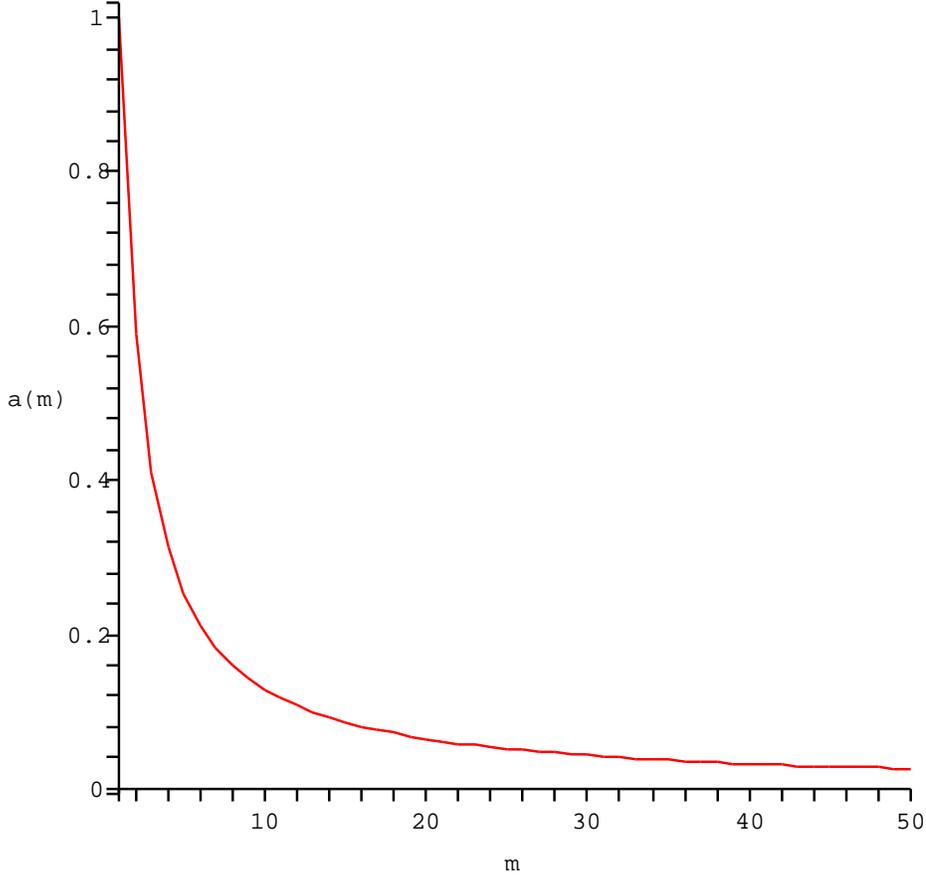}
	\caption{\label{good-a(m)} 
        Values of the sequence $a_m(\lambda
	= -1)$ where the initial values are chosen to satisfy $(4 \ln
	2 - 3) \alpha_0 = (2 \ln 2 - 1) \alpha_1$. This ensures the
	sequence converges to a constant value at large $m$.}
\end{figure}

With this in mind, we take as a particular example the choice $\lambda
= -1$; solving our differential equation for the generating function
$G(x)$ given in $(\ref{gen-ODE})$ with the condition $G(0)=\alpha_0$,
we obtain the solution
\begin{equation}
G(x) = \frac{\alpha_0 - (2 \alpha_0 + \alpha_1) x - (4 \alpha_0 + 2
\alpha_1) \ln (1-x)}{(1+x)^2}.
\end{equation}
In obtaining this solution from the differential equation
(\ref{gen-ODE}) for $G(x)$, we have used the relation (\ref{IC-rel})
between $a_0$ and the initial values $\alpha_0$ and $\alpha_1$. For
generic values $\alpha_0$ and $\alpha_1$, this function will have
singularities at $x= \pm 1$; to ensure that the singularity at $x=-1$
does not give rise to oscillatory behavior at large $m$, we require
that
\begin{equation}
\lim_{x\to -1} [(1 - x) (1 + x)^2 G(x)] = 0.
\end{equation}
When we solve this relation, we find that
\[
\alpha_1 = \biggl( \frac{4 \ln 2 - 3}{2 \ln 2 - 1} \biggr) \alpha_0
\]
which already implies that $(1-x)G(x)$ is regular at $x=-1$.
Notice that the $x=1$ singularity remains because of the $\ln(1 - x)$
term; however, $(1 - x) \ln(1 - x)$ is zero at $x=1$, so $(1-x)G(x)$
is regular. As discussed above, this is a ``good'' singularity where
the coefficients of the Taylor series go as $1/m$; this behavior in
$a_m$ is seen in Figure \ref{good-a(m)}. We remark here, however, that
only for negative $\lambda$ do we get the type of behavior seen above;
when $\lambda > 0$, the pole at $x=1$ is of higher order.

For arbitrary $\lambda$ we first solve the homogeneous equation
(\ref{gen-ODE}) for $a_0=0$ by
$G(x)=c(1+x)^{\lambda-1}(1-x)^{-\lambda-1}$. By varying the constant
$c$ we obtain the general solution to the inhomogeneous equation as
\begin{equation}
 G(x)=c(x)(1+x)^{\lambda-1}(1-x)^{-\lambda-1}
\end{equation}
with
\begin{equation}
 c(x)=a_0\int^x \left(\frac{1-t}{1+t}\right)^{\lambda}dt=
c_0-\frac{2^{\lambda}a_0}{\lambda-1} (1+x)^{1-\lambda}
\HGeo(1-\lambda,-\lambda;2-\lambda;(1+x)/2)
\end{equation}
for $\lambda\not=1$ in terms of the hypergeometric function
$\HGeo$. (If $\lambda=1$, the equation can be integrated in a 
manner similar to the $\lambda=-1$ case.) This gives
\begin{equation}
 G(x)=c_0(1+x)^{\lambda-1}(1-x)^{-\lambda-1}-
\frac{2^{\lambda}a_0}{\lambda-1} (1-x)^{-\lambda-1}
\HGeo(1-\lambda,-\lambda;2-\lambda;(1+x)/2)
\end{equation}
where only the first term determines the singularity structure at
$x=-1$ since the hypergeometric function $\HGeo(a,b;c;z)$ is regular
at $z=0$ taking the value $\HGeo(a,b;c;0)=1$ for all $a,b,c$. Thus,
the singularity at $x=-1$ can always be removed by choosing
$c_0=0$. Since
\begin{eqnarray*}
 a_1 &=& \alpha_0=G(0)=c_0-2^{\lambda}a_0/(\lambda-1)
 \HGeo(1-\lambda,-\lambda;2-\lambda;1/2)\\
 &=& c_0+a_0-\lambda
a_0(\psi(1/2-\lambda/2)-\psi(1-\lambda/2))
\end{eqnarray*}
with the digamma function $\psi(z)=d\Gamma(z)/dz$, this translates
to the condition
\begin{equation}
 a_1=a_0(1-\lambda(\psi(1/2-\lambda/2)-\psi(1-\lambda/2))\,.
\end{equation}
This expression is finite for all $\lambda$ which are not positive
integers since the digamma function is analytic except for simple
poles at $-z\in{\mathbb N}$. For $\lambda=-1$, for instance, we can
use $\psi(1)-\psi(3/2)=2\ln2-2$ and re-obtain the special case studied
before. 

At $x=1$, $\HGeo(a,b;c;(1+x)/2)$ has a branch point which is
logarithmic for $c-a-b\in{\mathbb Z}$ or $c-a-b\not\in{\mathbb
Q}$. Thus, $(1-x)G(x)$ always has a singularity at $x=1$, which for
positive $\lambda$ is enhanced by the factor
$(1-x)^{-\lambda-1}$. Thus, for $\lambda>0$ the sequence $\alpha_m$ is
unbounded.

\begin{figure}[htb]
	\includegraphics[width = 0.8\textwidth]{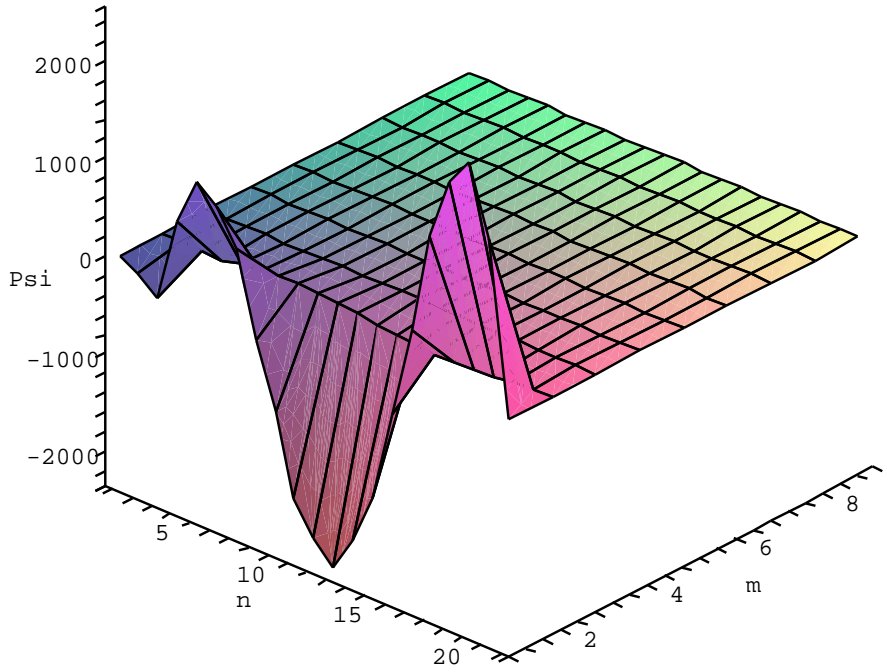}
	\caption{\label{sep-graph} Values of the sequence $t_{m, n}$
	where a linear combination of solutions $a_m b_n$ were chosen
	such that the wave function evolves into a Gaussian wave
	packet at large $n$.}
\end{figure}

\begin{figure}[htb]
	\includegraphics[width = 0.8\textwidth]{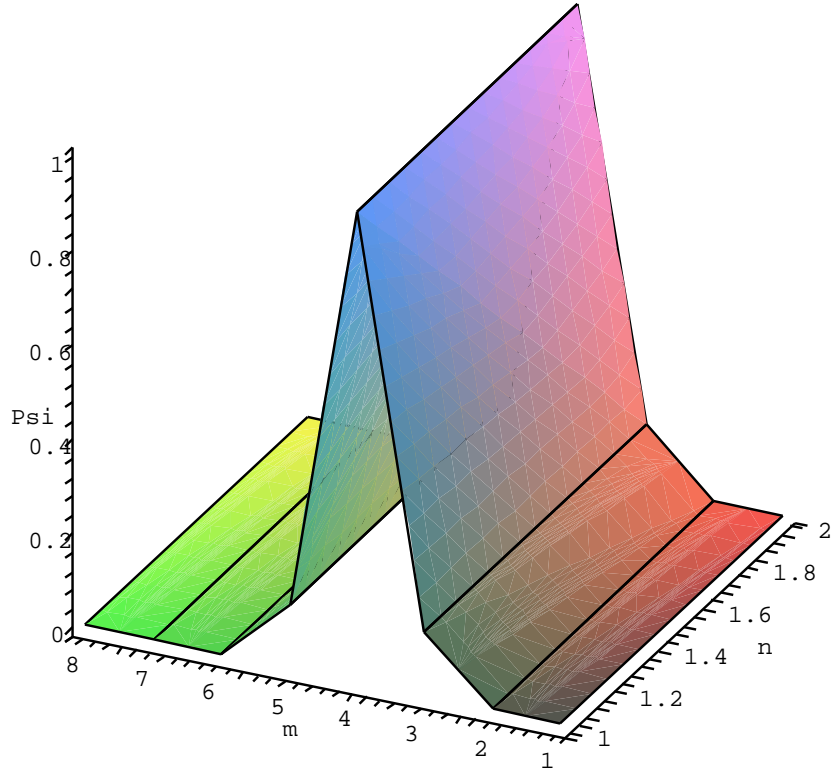}
	\caption{\label{final-sep} Close-up of the sequence given in
	Figure \ref{sep-graph}, showing the Gaussian wave packet at
	large values of $n$.}
\end{figure}

Now we turn to putting the $a_m$ and $b_n$ sequences back
together. For each $\lambda$, a pre-classical sequence $t_{m, n}$ --
when the boundary values have been appropriately tuned -- is
completely characterized by the choice of $\lambda$, with an
additional scaling factor. Because $b_0 = 0$ by the boundary
conditions on the sequence, changing $b_1$ will simply scale the
magnitude of the $b_n$ values. Similarly, pre-classicality requires a
relation between $a_0$ and $a_1$, so one is fixed by the choice of the
other, while varying this choice will again only scale the $a_m$
sequence. Since increasing $b_1$ by a constant factor and decreasing
both $a_0$ and $a_1$ by the same factor (to preserve their
relationship) leaves the $t_{m, n}$ sequence constant, only one of
these is independent. Thus, a general pre-classical solution $t_{m,
n}$ can be written as a sum $\sum_{\lambda \le 0} c(\lambda) t_{m, n}
(\lambda)$, with the coefficient $c(\lambda)$ used to scale the
individual terms of the sum. We can use linear combinations of these
to get a generic solution. As an example of this, we require that the
sequence $t_{m, n}$ match a Gaussian wave packet in the parameter $m$,
at large values of $n$ (see Figures
\ref{sep-graph} and \ref{final-sep}). One difficulty in doing this,
however, is apparent when looking at the $a_m$ graph, Figure
\ref{good-a(m)}; because the sequence becomes essentially constant for
large $m$, it is difficult to get the necessary discrimination in $m$
for the wave packet\footnote{Recall that the sequence $t_{m, n}$ was
the ``$m$ derivative'' of the original sequence ${\tilde t}_{m, n}$,
so one might wonder if these conclusions change when ${\tilde t}_{m,
n}$ is used. In fact, they remain much the same -- since $a_m$ gives
the change in ${\tilde t}_{m, n}$ with $m$, its decline to zero means
that ${\tilde t}_{m, n}$ is basically constant for large $m$. Thus,
again only at small $m$ is the spatial discrimination possible to
build a wave function with desired properties at large $n$.}.

\section{General solutions}
\label{full-sec}

In the previous section, we looked at separating out the $m$ and $n$
dependence into two distinct sequences, then using generating function
methods to analyze the $a_m$ sequence. Because this has reduced the
problem to finding the right initial values $a_0$ and $a_1$, there is
much to be said for the simplicity of this method. However, we can ask
whether the full recursion relation (\ref{diff-eqn}) is amenable to
study by generating functions. We shall see that this is feasible, and
provides a method to study other spacetimes where the recursion
relation may not be separable (which includes the Schwarzschild black
hole interior
\cite{BH}).

Starting again with the general difference equation
$(\ref{diff-eqn})$, we can work with a generating function $F(x, y)$,
which describes the sequence. Because occurrences of $n$ translate to
derivatives $\partial/\partial y$ of the generating function, the
function $d(n)$ on the surface presents a problem -- there would be a
non-polynomial function of derivatives acting on the generating
function. To deal with this, we write the sequence $t_{m, n}$ as
\begin{equation}
t_{m, n} = \sum_{k=0}^{\infty} t^{(k)}_{m, n},
\end{equation}
where the $t^{(k)}_{m, n}$ satisfy
\begin{subequations}
\begin{align}
\label{diff-eqn2}
\frac{1}{2n} [ t^{(0)}_{m+1, n} - t^{(0)}_{m-1, n}]  + \frac{2}{m} [
t^{(0)}_{m, n+1} - t^{(0)}_{m, n-1}] &= 0 \\
\frac{1}{2n} [ t^{(k)}_{m+1, n} - t^{(k)}_{m-1, n}]  + \frac{2}{m} [
t^{(k)}_{m, n+1} - t^{(k)}_{m, n-1}] &=
\left(\frac{1}{2n}-d(n)\right) (t^{(k-1)}_{m+1,n}-t^{(k-1)}_{m-1,n})\,.
\end{align}
\end{subequations}
Here, we are using the fact that $d(n) \simeq 1/2n$ to write the
sequence $t_{m, n}$ as a perturbation series.  In other words, we
choose the leading order sequence $t^{(0)}_{m, n}$ to satisfy a
recursion relation with coefficients polynomial in the parameters $m$
and $n$, such that a partial differential equation can be found that
is equivalent to the $t^{(0)}_{m, n}$ relation
(\ref{diff-eqn2}). Boundary or initial values would then be set such
that $t^{(0)}$ is pre-classical, i.e.\ by removing singularities of
the generating function. For higher terms $t^{(k)}$, $k\geq 1$, we can
choose vanishing boundary and initial values. Each such contribution
by itself would not be pre-classical, but its oscillations would
contribute only small perturbations to $t^{(0)}$. Indeed, because of
the small difference between $1/2n$ and $d(n)$, the right hand side of
the equations for $t^{(k)}_{m,n}$, $k>0$ is suppressed by a factor of
$n^{-3}$ compared to the left hand side. It is thus consistent with
pre-classicality to use the relation for $t^{(0)}$ in order to fix
boundary values as before, and then solve the relations for higher
$t^{(k)}$ with boundary and initial values zero. With a vanishing
right hand side the solution would be $t^{(k)}_{m,n}=0$ for $k\geq 1$,
and the small right hand side will lead to small values for the higher
terms which become negligible compared to the $t^{(0)}_{m, n}$
values. Because of this, for the rest of the paper we focus on solving
for the sequence $t^{(0)}_{m, n}$; obtaining finer accuracy for small
$m, n$ would be possible using higher $t^{(k)}_{m, n}$

Working with the $t^{(0)}_{m, n}$ sequence, we look for a partial
differential equation for the generating function
\begin{equation}
F(x, y) = \sum_{m=0} ^\infty \sum_{n=0} ^\infty t^{(0)}_{m, n} x^m y^n.
\end{equation}
Multiplying the recursion relation (\ref{diff-eqn2}) for $t^{(0)}_{m,
n}$ by $(1/2) mn x^{m-1} y^{n-1}$ and summing over all values it is
valid gives
\begin{equation}
 \sum_{m=0} ^\infty \sum_{n=0} ^\infty \biggl [ \frac{1}{4} (m+1)
 (t^{(0)}_{m+2, n+1} - t^{(0)}_{m, n+1} ) + (n+1) (t^{(0)}_{m+1, n+2}
 - t^{(0)}_{m+1, n}) \biggr ] x^m y^n = 0,
\end{equation}
and we arrive at an equation for $F(x, y)$ which is similar in form to
the separable case, i.e. there are singularities at $x, y = 0$. To
avoid this, we define a new function $G(x, y)$ as
\begin{equation}
G(x, y) = \frac{1}{xy} \biggl[ F(x, y) - \sum_{n=1} ^\infty t^{(0)}_{0, n}
y^n \biggr] = \frac{[ F(x, y) - y A(y) ]}{xy},
\end{equation}
where
\begin{equation}
A(y) = \sum_{n=1} ^\infty t^{(0)}_{0, n} y^{n - 1}
\end{equation}
is related to the generating function of the $m=0$ boundary for the
original $t^{(0)}_{m, n}$ sequence (the definition of $A(y)$ chosen here 
is to simplify the form of our eventual differential equation for $G(x, y)$). 
This sum begins at $n=1$, since the boundary conditions imply $t^{(0)}_{0, 0} = 0$. 
This gives a ``shifted'' sequence, since
\begin{equation}
F(x, y) = xy G(x, y) + y A(y).
\end{equation}
Note the similarity between this relation between $F(x, y)$ and $G(x,
y)$, and the analogous one between $F(x)$ and $G(x)$ in the separable
case. Again, the coefficients are shifted by one, now in each of the
parameters; if
\begin{equation}
G(x, y) = \sum_{m=0} ^\infty \sum_{n=0} ^\infty \rho_{m, n} x^m y^n
\end{equation}
then
\begin{equation}
t^{(0)}_{m, n} = \rho_{m-1, n-1}.
\end{equation}
This gives us the PDE
\begin{equation}
 \frac{1}{4} \frac{\partial}{\partial x} [(1 - x^2) G(x, y)] +
 \frac{\partial}{\partial y} [(1 - y^2) G(x, y)] = A(y).
\end{equation}
Again, we will have singularities at both $x = \pm 1$ and $y = \pm 1$.

Our discussion of the singularities in Section \ref{sep-sec} for
separable solutions carries over here similarly for functions of two
variables. First, we want the functions $(1-x) G(x, y)$ and $(1-y)G(x,
y)$ both to be finite for all values of $x$ and $y$, respectively, to
ensure that at large $m$ or $n$, the sequence is either constant or
smoothly changing. Putting these together means that $(1-x)(1-y) G(x,
y)$ has to be finite for all $x, y$. As with the separable solutions,
the only problem with this might be at $x=-1$ and $y=-1$. One
condition for $(1-x)(1-y) G(x, y)$ to be finite along these two lines in
the $x-y$ plane is that the function
\begin{equation}
\label{G-H-eqn}
H(x, y) = (1 - x^2)(1 - y^2) G(x, y)
\end{equation}
satisfies $H(-1, y) = H(x, -1) = 0$. When we solve for $G(x, y)$ in
terms of $H(x, y)$ in the above relation (\ref{G-H-eqn}), we find the
new equation
\begin{equation}
 \frac{1}{4} (1 - x^2) \frac{\partial H(x, y)}{\partial x} + (1 - y^2)
 \frac{\partial H(x, y)}{\partial y} = (1 - x^2) (1 - y^2) A(y).
\end{equation}
So, given our boundary conditions on $H(x, y)$ mentioned above, and
any choice for $A(y)$, we can solve this equation numerically.

\begin{figure}
	\includegraphics[width = 0.8\textwidth]{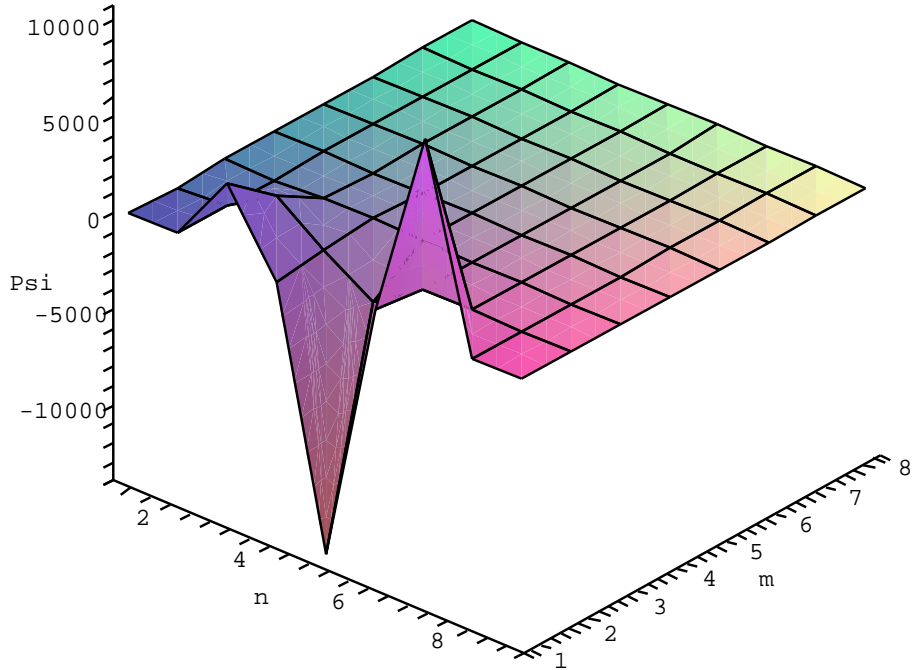}
	\caption{\label{spectral} Values of the sequence $r_{m, n}$
	obtained from the Taylor expansion of the generating function
	$G(x, y)$. Notice the similarity in shape with Figure
	\ref{sep-graph} obtained by combining the one-parameter
	sequences $a_m$ and $b_n$.}
\end{figure}

This is exactly how we would go about finding the sequence if there
was a desired sequence of values $t_{0, n}$, which is what $A(y)$
specifies. However, more likely one wants a particular limiting case
for the sequence $t_{m, n}$ for large $n$, for example. Here, the
properties of the functions $G(x, y)$ and $H(x, y)$ come into
play. Because it is assumed that $(1-x)(1-y) G(x, y)$ is finite, then
$G(x, y)$ itself can have simple poles at $x=1$ and $y=1$. For the
cases mentioned in Section \ref{sep-sec} -- the functions $(1-x)^{-1}$
and $\ln(x-1)$ -- the coefficients in the Taylor series converge to
constant values (unity in the first sequence, zero in the second). So,
if one looks at the expansion in $y$, the coefficients of $G(x, y)$
will assume some constant profile in $x$ for large powers $y^n$; this
profile will be determined by, e.g., the residue of the function $G(x,
y)$ at a simple pole $y=1$. Up to a constant factor, this is simply
the function $H(x, 1)$; along the lines $x=1$ and $y=1$, the function
$H(x, y)$ gives us information about the asymptotic behavior of the
sequence $t_{m, n}$. In fact, $H(x, 1)$ is simply the generating
function reflecting how $t_{m, n}$ behaves at large $n$, and similarly
for $H(1, y)$. Thus, to assemble the desired profile at, say, large
$n$, one requires that $H(x, 1)$ be a certain function, treating the
$y^n$ coefficients of the source term $A(y)$ as unknowns to be solved
for. This can be done either numerically or by using pseudo-spectral
methods; the form of the partial differential equation makes it
particularly amenable to the use of Chebyshev polynomials. An example
of this is shown in Figure \ref{spectral}; notice that this has the
same basic shape as the sequence assembled out of the separable
solutions in Section \ref{sep-sec}.

\section{Conclusions}

Loop quantum cosmology has discrete evolution equations of higher
order than the Wheeler--DeWitt differential equation which is obtained
in a continuum limit. There are thus additional independent solutions
which display an alternating behavior. Those solutions are not
necessarily unphysical and can in fact contain important information
about quantum effects. Nevertheless, for studying the semiclassical
limit of those models it is important to have control on the subset of
solutions for which the small-scale oscillations are suppressed. In
particular, one needs to find selection criteria for this subset,
usually by specifying initial or boundary values. Choosing those
special values, if possible at all, is thus not done to achieve a 
particular physical consequence and does not amount to fine-tuning. As
demonstrated here for the Bianchi I LRS model, appropriate solutions
can be extracted analytically by employing generating function
techniques. It is emphasized here that these methods will work with
other partial difference equations, so they can be applied to any LQC
model.

For this particular model we can apply the results to discuss the
initial value problem. In general, one would need separated solutions
for all values of the separation parameter in order to construct a
given initial wave packet at large $n$, far away from the classical
singularity at $n=0$. The Wheeler--DeWitt equation with DeWitt's
initial conditions does not allow sufficiently many solutions and thus
presents an ill-posed initial value problem. For the difference
equation of loop quantum cosmology, the initial value problem is
well-posed in the sense that there are always non-trivial
solutions. However, in the semiclassical limit one also has to find enough
solutions which do not oscillate strongly at small scales. One may
then speculate that analogously to the ill-posed Wheeler--DeWitt
problem there are not sufficiently many pre-classical
solutions. Another system which is ill-posed from the Wheeler--DeWitt
point of view, an isotropic model with a free, massless scalar has
been studied in \cite{Scalar}. There it turned out that the loop
problem is well-posed and still allows enough pre-classical
solutions. However, the situation there was different in that the
matter term was responsible for the ill-posedness of the continuum
formulation. The Bianchi I LRS model thus presents a qualitatively
different system to study well-posedness, which can be done in detail
with the methods developed in this paper.

What we have seen is that for negative separation parameter $\lambda$
we can find pre-classical solutions. The problems of the
Wheeler--DeWitt quantization at the boundary $p_A=0$ do not occur in
the discrete formulation and pre-classicality puts strong restrictions
on the sequence $a_m$ fixing its boundary values. The sequence $b_n$
has only small oscillations for $\lambda<0$ such that one can choose a
pre-classical solution.

For positive $\lambda$, on the other hand, the situation is
different. Now, the $a_m$ sequence is not problematic, and for $b_n$
one would have to choose special initial values in order to guarantee
pre-classicality. However, since $b_0=0$ is required by the difference
equation, this option is not available and $b_n$ will always be
alternating for positive $\lambda$. Thus, $b_n$ does not have a
continuum limit but $(-1)^nb_n$ does. However, the corresponding
continuum limit of the difference equation acting on $(-1)^n a_mb_n$
has an additional relative minus sign from the $(-1)^n$ such that one
obtains a different Wheeler--DeWitt equation: Introducing
$T_{m,n}:=(-1)^nt_{m,n}$ into the difference equation (\ref{diff-eqn}) 
for $t_{m, n}$, we obtain
\begin{eqnarray*}
 && (-1)^n d(n) (T_{m+1,n}-T_{m-1,n}) +
 2d_2(m)((-1)^{n+1}T_{m,n+1}-(-1)^{n-1}T_{m,n-1})\\
 &=& (-1)^n [d(n) (T_{m+1,n}-T_{m-1,n}) -
 2d_2(m)(T_{m,n+1}-T_{m,n-1})]=0
\end{eqnarray*}
where the sign of the second term is switched. This is an example
of different continuum limits obtained from one and the same
difference equation, which, from the point of view of the discrete
formulation, can be interpreted as a duality between different
continuum formulations \cite{cosmoIV}.

In order to construct semiclassical wave packets at large $n$, one can
try to use only negative $\lambda$ in order to avoid
oscillations. However, the resulting solutions are mostly concentrated
at small values of $m$ since the pre-classical separable solutions
$a_m$ decrease with $m$. One can also see that using only negative
$\lambda$ is not enough by observing that the Wheeler--DeWitt
solutions $p_A^{\lambda}$ for positive integer $\lambda$ are just what
one needs for a Taylor expansion from which one could construct any
analytical initial wave packet. Using these $\lambda$ in the discrete
case requires to have alternating $b_n$, which implies that such wave
packets do not follow the classical trajectory since they are
solutions to the Wheeler-DeWitt equation with a sign flip.

Thus, in this model the discrete analog of DeWitt's initial condition
may be problematic. It is possible to avoid DeWitt's initial condition
by using a symmetric version of the constraint (which is non-singular
if the constraint is symmetrized after multiplying with ${\rm sgn}(n)$
\cite{IsoCosmo}). In this way detailed studies of models can teach
lessons for the full theory since the same ordering should be used in
all cases. With this symmetric ordering, $b_0$ would be free such that
it can be fixed as done for the $a_m$ sequence if one requires a
pre-classical solution. However, due to the symmetrization of the
constraint operator the difference equation will change and be more
complicated; moreover, one loses conditions on the wave function.
Even with the non-symmetric constraint used here, it follows from the
results of this paper that, with the condition of pre-classicality
in addition to dynamical initial conditions, the wave function will not
be unique.

Alternatively, one can take the lack of pre-classical solutions
seriously for this model. If we only look at large values of $m$ and
$n$, i.e.\ only at the semiclassical regime, there is no problem at
all and we can construct any initial state we like. Most of those
states would not solve the initial conditions imposed by the quantum
constraint at the classical singularity and thus have to be
discarded. At this point, however, we already use information from the
quantum theory: the behavior of the wave function at the classical
singularity is invoked. Thus, the properties of wave functions
discussed here can be interpreted as a quantum effect.

\acknowledgments

GK is grateful for research support from the University of
Massachusetts at Dartmouth, as well as NSF grant PHY-0140236. The
authors also appreciate helpful discussions with Rodolfo Gambini,
Robert Israel and Jorge Pullin.

\end{document}